\documentclass[twocolumn]{aastex631}
\usepackage{amsmath}
\usepackage{mathrsfs}
\received{...}
\revised{...}
\accepted{...}
\published{...}

\begin{document}

\title{Upper Limit of Sound Speed in Nuclear Matter: A Harmonious Interplay of Transport Calculation and Perturbative Quantum Chromodynamic Constraint}

\author[0000-0001-9120-7733]{Shao-Peng Tang}
\affiliation{Key Laboratory of Dark Matter and Space Astronomy, Purple Mountain Observatory, Chinese Academy of Sciences, Nanjing 210033, China}
\author[0000-0002-7505-7795]{Yong-Jia Huang}
\affiliation{Key Laboratory of Dark Matter and Space Astronomy, Purple Mountain Observatory, Chinese Academy of Sciences, Nanjing 210033, China}
\affiliation{RIKEN Interdisciplinary Theoretical and Mathematical Sciences Program (iTHEMS), RIKEN, Wako 351-0198, Japan}
\author[0000-0001-9034-0866]{Ming-Zhe Han}
\affiliation{Key Laboratory of Dark Matter and Space Astronomy, Purple Mountain Observatory, Chinese Academy of Sciences, Nanjing 210033, China}
\affiliation{Max-Planck-Institut f\"ur Gravitationsphysik (Albert-Einstein-Institut), Am M\"uhlenberg 1, D-14476 Potsdam-Golm, Germany}
\author[0000-0002-8966-6911]{Yi-Zhong Fan}
\affiliation{Key Laboratory of Dark Matter and Space Astronomy, Purple Mountain Observatory, Chinese Academy of Sciences, Nanjing 210033, China}
\affiliation{School of Astronomy and Space Science, University of Science and Technology of China, Hefei, Anhui 230026, China}
\correspondingauthor{Yi-Zhong Fan}
\email{yzfan@pmo.ac.cn}

\begin{abstract}
Very recently, it has been shown that there is an upper bound on the squared sound speed of nuclear matter from the transport, which reads $c_{\rm s}^2 \leq 0.781$. 
In this work, we demonstrate that this upper bound is corroborated by the reconstructed equation of state (EOS; modeled with a nonparametric method) for ultradense matter.
The reconstruction integrates multimessenger observation for neutron stars, in particular, the latest radius measurements for PSR J0437$\textendash$4715 ($11.36^{+0.95}_{-0.63}$ km), PSR J0030+0451 ($11.71^{+0.88}_{-0.83}$ km, in the ST+PDT model), and PSR J0740+6620 ($12.49^{+1.28}_{-0.88}$ km) by NICER have been adopted.
The result shows in all cases, the $c_{\rm s}^2 \leq 0.781$ upper limit for EOS will naturally yield the properties of matter near the center of the massive neutron star consistent with the causality-driven constraint from pQCD, where, in practice, the density in implementing the pQCD likelihood ($n_{\rm L}$) is applied at $10n_s$ (where $n_s$ is the nuclear saturation density). We also note that there is a strong correlation for the maximum $c_s^2$ with $n_{\rm L}$, and $c_{\rm s}^2 \leq 0.781$ is somehow violated when $n_{\rm L} = n_{\rm c,TOV}$. The result indicates that a higher $n_{\rm L}$, even considering the uncertainties from statistics, is more natural.
Moreover, the remarkable agreement between the outcomes derived from these two distinct and independent constraints (i.e., the transport calculation and pQCD boundary) lends strong support to their validity.
In addition, the latest joint constraint for $R_{1.4}$, $R_{2.0}$, $R_{1.4}-R_{2.0}$, and $M_{\rm TOV}$ are $11.94_{-0.68}^{+0.77}$ km, $11.99_{-0.67}^{+0.88}$ km, $-0.1_{-0.27}^{+0.42}$ km, and $2.24_{-0.10}^{+0.13}M_\odot$ (at $90\%$ credible level), respectively.
\end{abstract}

\section{Introduction} \label{sec:intro}
The observation of neutron stars (NSs) with masses surpassing $2M_\odot$ \citep{2013Sci...340..448A, 2021ApJ...915L..12F} implies that the speed of sound must exceed the conformal limit at certain densities \citep{2015PhRvL.114c1103B} and the speed of sound (as a function of density) within the NS EOS is believed to be nonmonotonic, characterized by the presence of at least one peak \citep{2020NatPh..16..907A, 2022ApJ...939L..34A, 2023ApJ...950..107G, 2023SciBu..68..913H, 2023ApJ...949...11J, 2024PhRvC.109f5803Y}. This has raised the general theoretical question: How high can the speed of sound be within physically plausible systems or theories?
In a recent study, \citet{2024arXiv240214085H} summarized the transport coefficients for nuclear matter in many theories \citep{2001PhRvL..87h1601P, 2008JHEP...04..100B, 2007PhRvL..99q2301R, 2012JHEP...11..148M, 2003JHEP...05..051A, 2017arXiv171205815R} and found they would finally yield an upper limit of $c_s^2\leq 0.781$ for all known systems.
It is, however, noted that in principle, this bound could be surpassed in some ideal systems \citep{2024JHEP...06..171M}.
In this work, with the information from various directions, we show that the constraint for nuclear matter sound speed will naturally lead to the properties of matter near the center of the most massive NS well consistent with the causality-driven constraint from the perturbative quantum chromodynamic (pQCD).

Advancements in pQCD calculations have garnered significant interest due to their contributions in bounding the NS EOS \citep{2020NatPh..16..907A, 2022ApJ...939L..34A, 2022PhRvL.128t2701K, 2022arXiv221111414K, 2023ApJ...950..107G, 2023SciBu..68..913H, 2023PhRvD.108i4014B, 2023PhRvD.107a4011B, 2023arXiv230902345M, 2023PhRvC.107e2801S, 2023arXiv230711125Z, 2024PhRvD.109i4030K, 2024PhRvD.109d3052F, 2024AcPPB..55....1V}.
Given the challenge of inferring the state of matter in the vicinity of the NS core from observable NS properties, the incorporation of insights from pQCD is essential for achieving a more rigorous constraint on the EOS.
Although the perturbative calculation is only valid in ultrahigh density, such boundary conditions could be extended to lower chemical potential based on thermodynamic stability and causality \citep{2022PhRvL.128t2701K, 2023ApJ...950..107G}. As a result, soft cores are generally required in massive NS. However, it is still in debate; one reason is whether the soft core being required somehow depends upon the density to implement the causality-driven constraint from pQCD. Various studies have examined the density at which the pQCD likelihood \citep{2023ApJ...950..107G} should be implemented, with no consensus reached thus far \citep{2023PhRvC.107e2801S, 2023ApJ...950..107G, 2023PhRvD.108d3013E,  2023PhRvD.108i4014B, 2023PhRvD.107a4011B, 2023arXiv230902345M, 2024PhRvD.109d3052F}.

In this work, we study the NS EOS and its sound speed by using the multimessenger observations and the theoretical constraint. We incorporate all of the information with Bayesian inference, while the Gaussian-process EOS generator guarantees the method is nonparametric. We examine how the upper bound of $c_s^2$ affects the EOS of ultradense matter and demonstrate that the maximal values of $c_s^2$, derived from multimessenger NS observations and pQCD likelihood \citep{2022PhRvL.128t2701K, 2023ApJ...950..107G} that were calculated at $n_{\rm L}=10n_{\rm s}$ (where $n_{\rm s}$ is the nuclear saturation density), are in good agreement with the theoretical upper bound of $c_s^2 \leq 0.781$.
However, this upper bound is partially violated when the pQCD likelihood is used at $n_{\rm L}=n_{\rm c,TOV}$, the central density of the maximum mass NS configuration.
These results provide support for the application of pQCD likelihood at densities surpassing those found within NSs, potentially leading to a more precise understanding of the EOS for NSs and the behavior of matter under extreme conditions.
Moreover, by applying the transport upper bound on the constructed EOS but without pQCD information, we can corroborate many projections predicated on causality-driven constraint (applied at $10n_{\rm s}$) from pQCD. The concordance observed between these different physical models mutually reinforces their validity.

\section{Methods}\label{sec:methods}
Building upon the methodology delineated in our previous works \citep{2024PhRvD.109h3037T} (see also \citealt{2023ApJ...950..107G} and \citealt{2024PhRvD.109d3052F}), we implement the Gaussian-process method \citep{2019PhRvD..99h4049L, 2020PhRvD.101f3007E, 2020PhRvD.101l3007L} to generate an ensemble of EOSs.
The posterior distribution of the EOS is constructed by selecting samples based on their likelihoods, denoted as $\mathcal{L} = \mathcal{L}_{\rm GW} \times \mathcal{L}_{\rm NICER}  \times \mathcal{L}_{\rm M_{\rm max}} \times \mathcal{L}_{\rm pQCD}$.
Here, $\mathcal{L}_{\rm GW}$ quantifies the likelihood associated with the mass-tidal deformability measurements from GW170817 \citep{2018PhRvL.121p1101A, 2019PhRvX...9a1001A}, and $\mathcal{L}_{\rm NICER}$ corresponds to the likelihood of NICER observations, specifically the latest mass and/or radius measurements for PSR J0030+0451 
(i.e., the ST+PDT model results reported in \citet{2024ApJ...961...62V}, since the PDT-U model is disfavored \citep{2024ApJ...966...98L}), PSR J0740+6620 \citep{2024arXiv240614466S, 2024arXiv240614467D}, and PSR J0437$\textendash$4715 ($R_{1.418M_\odot} = 11.36_{-0.63}^{+0.95}$ km at the $68.3\%$ credible level\footnote{Intriguingly, this radius measurement is well consistent with that of PSR J0030+0451 in the ST+PDT model \citep{2024ApJ...961...62V}.} \citep{2024ApJ...971L..20C}).
The term $\mathcal{L}_{\rm M{\rm max}}$ is the marginalized posterior distribution of the maximum mass cutoff for NSs, as adopted from \citet{2024PhRvD.109d3052F}.
Lastly, $\mathcal{L}_{\rm pQCD}=\mathcal{P}(n_{\rm L}, \varepsilon(n_{\rm L},{\rm EOS}), p(n_{\rm L}, {\rm EOS}))$ represents the likelihood of the pQCD constraints at a density of $n_{\rm L}$ \citep{2022PhRvL.128t2701K, 2023ApJ...950..107G}, where $\varepsilon$ and $p$ denote the energy density and pressure, respectively.
In this study, we examine the implications of applying the pQCD likelihood in two distinct scenarios: one where the likelihood is applied at the core density corresponding to the maximum mass of nonrotating NSs and another at a density of $n_{\rm L}=10n_{\rm s}$.
The extension of pQCD to lower densities, such as $10n_{\rm s}$, involves the consideration of two distinct upper limits for the squared speed of sound. The first upper limit is based on the principle of causality and is represented as pQCD($10n_{\rm s}, c_s^2 \leq 1$), while the second is derived from transport and is denoted as pQCD($10n_{\rm s}, c_s^2 \leq 0.781$).
Furthermore, we explore the implications of applying a $c_s^2$ upper limit that is applicable to the entire EOS, extending to densities well beyond that of an NS, and is specified as $c_s^2(n \leq 40n_{\rm s}) \leq 0.781$. This investigation is conducted without including the pQCD likelihood term.

\section{Results} \label{sec:results}
\begin{figure}
    \centering
    \includegraphics[width=0.49\textwidth]{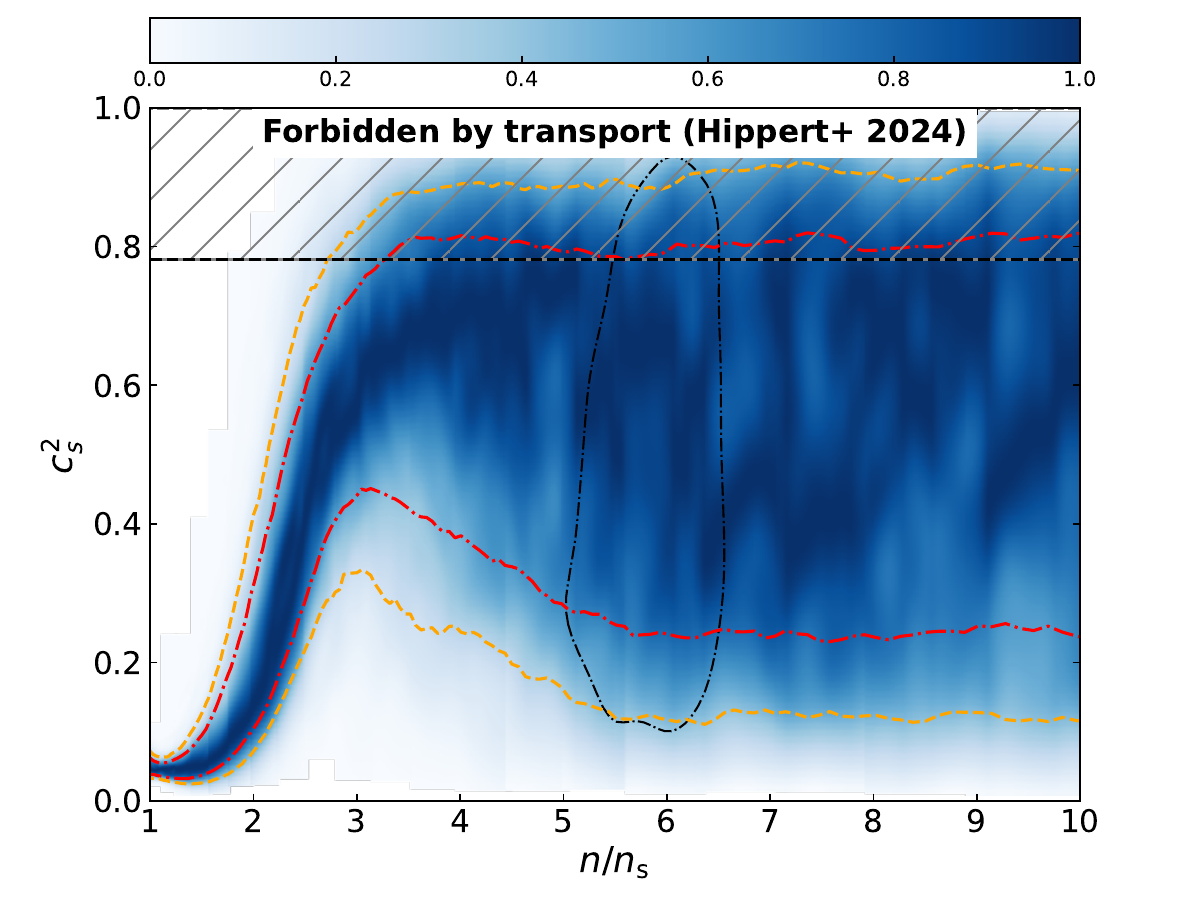}
    \includegraphics[width=0.49\textwidth]{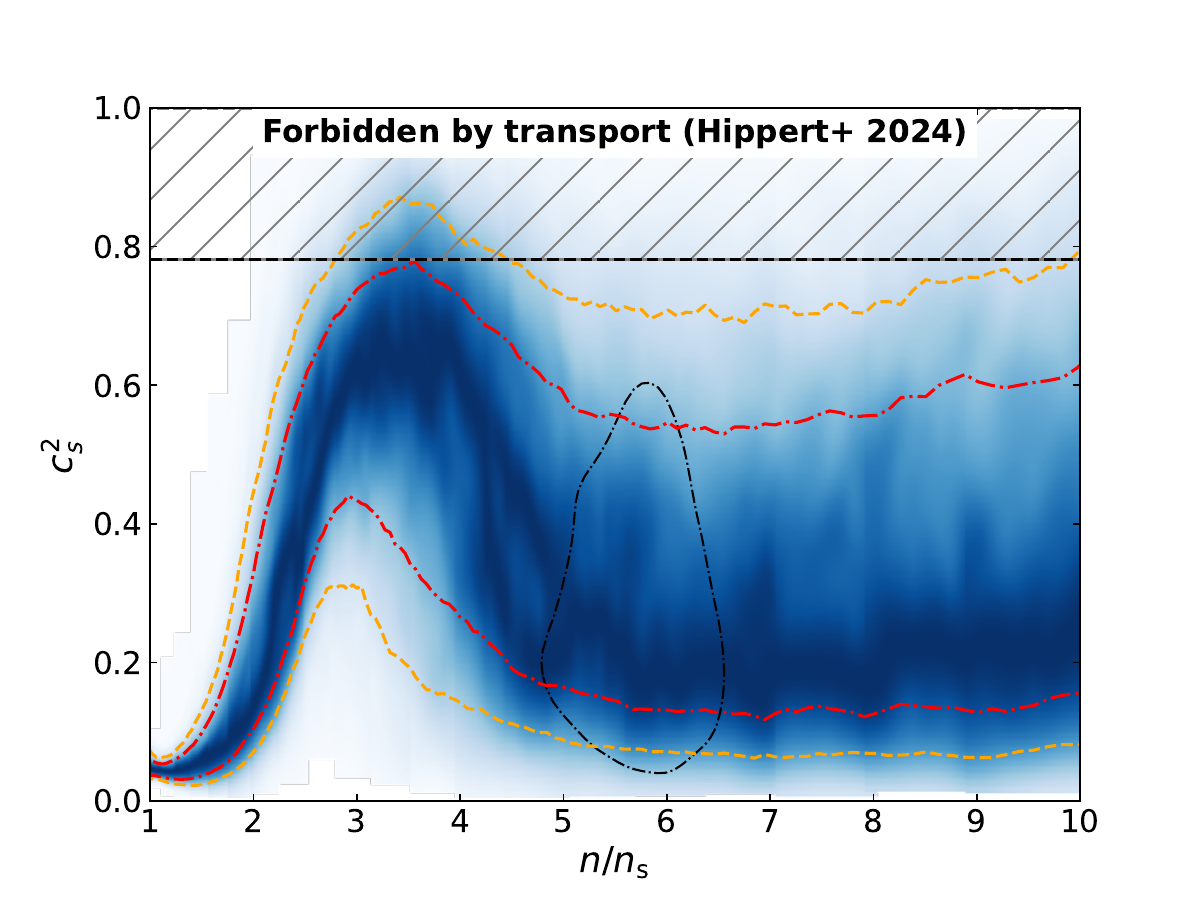}
    \caption{Density distribution of the sound speed squared ($c_s^2$) as a function of the number density $n$ normalized to $n_{\rm s}$. Top panel: this result is derived using $n_{\rm L}=n_{\rm c,TOV}$. The red dashed-dotted and orange dashed curves represent the $68.3\%$ and $90\%$ confidence regions, respectively. The transport bound $c_s^2\leq0.781$ \citep{2024arXiv240214085H} is delineated by a hatched pattern and labeled as ``forbidden by transport." Additionally, the black dashed-dotted line indicates the $68.3\%$ contour for $n_{\rm c,TOV}$ and its associated values of $c_s^2$. Bottom panel: the same as top panel but derived using $n_{\rm L}=10n_{\rm s}$. The results in both panels are obtained with $c_s^2 \leq 1$ extension of pQCD to lower densities.}
    \label{fig:n-cs2-1}
    \hfill
\end{figure}
As depicted in the top panel of Figure~\ref{fig:n-cs2-1}, the square of the speed of sound, $c_s^2(n)$, is reconstructed utilizing multimessenger NS observations in conjunction with the likelihood of pQCD at a density of $n_{\rm L} = n_{\rm c,TOV}$.
We find that the reconstructed $c_s^2$ exhibits subconformality ($c_s^2 \leq 1/3$) at low densities and there is a rapid increase in the value of $c_s^2$ as the density increases from $1.5n_{\rm s}$ to $3n_{\rm s}$.
Beyond $3n_{\rm s}$, the $90\%$ upper limit for $c_s^2$ surpasses the transport bound of $c_s^2 \leq 0.781$, suggesting that either the pQCD likelihood applied at $n_{\rm L} = n_{\rm c,TOV}$ is not supported or that the matter composition beyond $3.5n_{\rm s}$ potentially deviates from `ordinary' nuclear matter.
At densities above $5n_{\rm s}$, the constraints on $c_s^2$ become significantly less stringent due to the reduced informativeness of multimessenger data and the very weak influence of the pQCD likelihood with $n_{\rm L} = n_{\rm c,TOV}$.

Upon implementing pQCD likelihood at a density of $n_{\rm L} = 10n_{\rm s}$, the EOS within the density range of $1.5n_{\rm s}$ to $3n_{\rm s}$ exhibits a similar stiffness to that observed when $n_{\rm L} = n_{\rm c,TOV}$.
However, at densities exceeding $3.5n_{\rm s}$, the squared speed of sound decreases rapidly, exhibiting a pronounced peak at $3.5n_{\rm s}$, with a magnitude in agreement with the transport bound, as depicted in the bottom panel of Figure~\ref{fig:n-cs2-1}.
This phenomenon is primarily attributed to the stiffening of the EOS required to support NSs with masses surpassing $2M_\odot$ \citep{2013Sci...340..448A, 2021ApJ...915L..12F} and the softening of the EOS by pQCD likelihood.
In contrast to the $n_{\rm L} = n_{\rm c,TOV}$ scenario, the sound speed at $n_{\rm c,TOV}$ is more likely to fall below the conformal limit.
This suggests that the core of a maximum mass NS may contain matter that deviates from pure hadronic compositions.
Across higher-density regions, the $c_s^2$ values roughly distributed around approximately $0.2$, a result attributed to the pQCD likelihood applied at large density (e.g., $n_{\rm L}\geq 10n_{\rm s}$).
We also observe that the majority of the $c_s^2(n)$ results across the entire density range adhere to the established upper limit of $c_s^2$ as informed by nuclear matter transport calculations.
The extension of pQCD to lower densities is contingent upon the upper bound of the speed of sound squared. We have verified that employing an upper limit of $c_s^2 \leq 0.781$ does not affect the outcomes derived from causal extrapolation. For additional details, please refer to the Appendix. 

\begin{figure}
    \centering
    \includegraphics[width=0.49\textwidth]{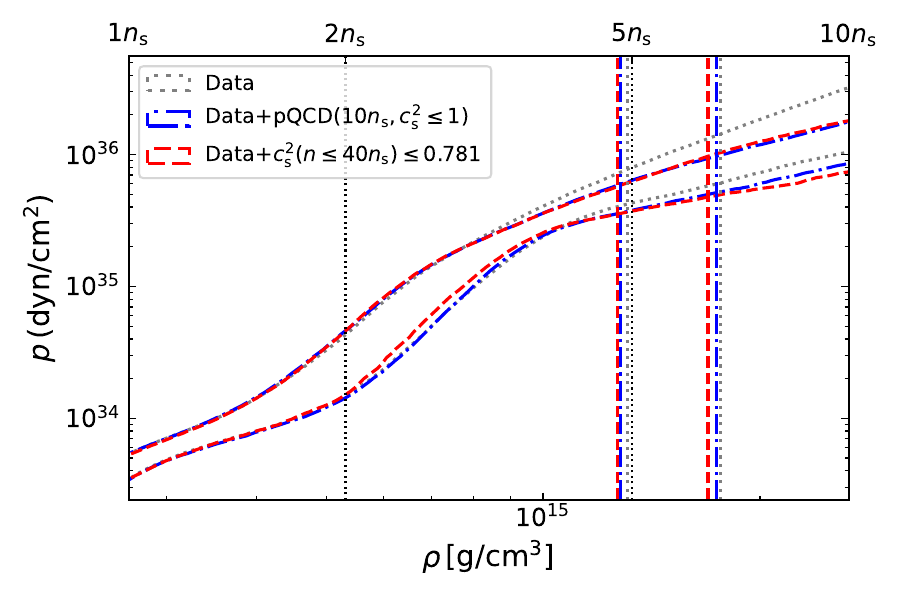}
    \caption{Reconstructed $90\%$ credible regions for the pressure versus rest-mass density relationship. The gray dotted lines represent the results derived from multimessenger NS observations. The blue and red lines correspond to results obtained when an additional likelihood from pQCD is applied at a density of $10n_{\rm s}$ and when incorporating additional transport upper bound of $c_s^2\leq0.781$, respectively.}
    \label{fig:rho-p}
    \hfill
\end{figure}
Below, we assess the consistency between the transport calculations and the pQCD likelihood applied at densities of $n_{\rm L}=10n_{\rm s}$. We establish an upper bound for the speed of sound squared, $c_s^2$, within the proposed EOS based on transport calculation \citep{2024arXiv240214085H}. These bounded-EOS results are then compared with the results derived from pQCD likelihood applied at $n_{\rm L}=10n_{\rm s}$, which incorporate causality extrapolation. As shown in Figure~\ref{fig:rho-p}, there is a remarkable agreement between these two approaches (refer to Figure~\ref{fig:nnctov-p} in the Appendix for results normalized to $n_{\rm c,TOV}$). Additionally, both methodologies indicate a softening of the EOS at higher densities compared to that  obtained with only multimessenger observations of NSs.
\begin{figure}
    \centering
    \includegraphics[width=0.49\textwidth]{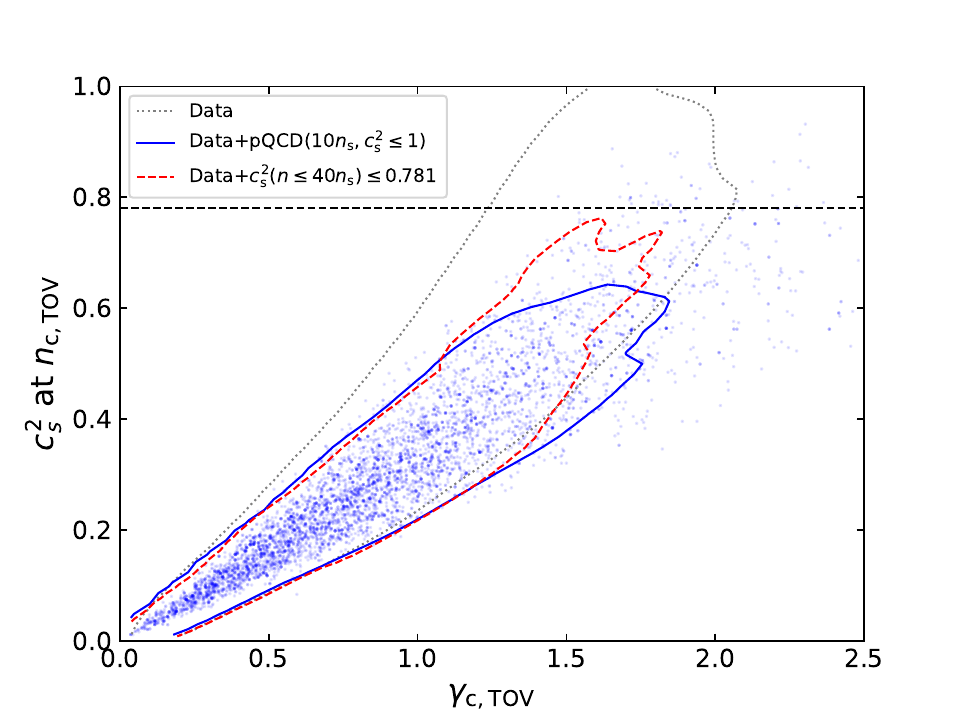}
    \caption{The properties ($c_s^2$ vs. $\gamma$) of the matter at the center of the most massive nonrotating NS. The gray dotted lines represent the results derived from multimessenger NS observations. The blue and red lines correspond to results ($90\%$ credibility) obtained for the pQCD likelihood applied at $n_{\rm L}=10n_{\rm s}$ and for the additional transport upper bound of $c_s^2\leq0.781$.}
    \label{fig:cs2-gamma}
    \hfill
\end{figure}
The central state of the most massive nonrotating NS, characterized by the squared speed of sound ($c_s^2$) and the polytropic index ($\gamma_{\rm c,TOV}$), demonstrates significant uniformity (see Figure~\ref{fig:cs2-gamma}).
\begin{figure}
    \centering
    \includegraphics[width=0.49\textwidth]{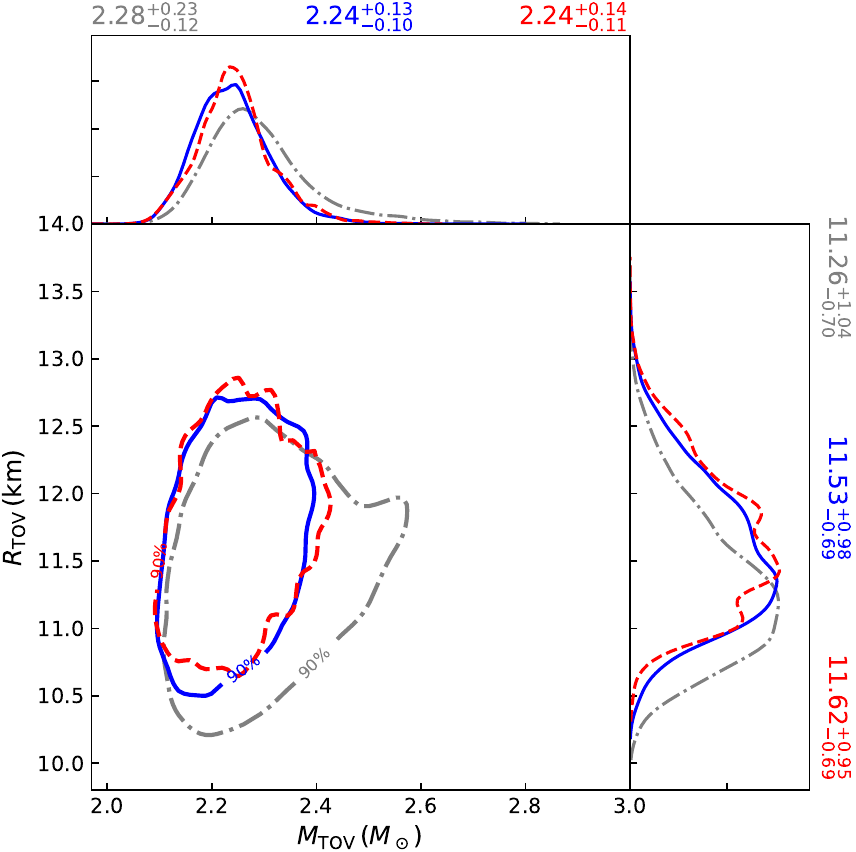}
    \caption{Posterior distributions of $M_{\rm TOV}$ and $R_{\rm TOV}$ ($90\%$ credibility). The gray dashed-dotted lines represent the results derived from multimessenger NS observations. The blue and red lines correspond to results obtained when an additional likelihood from pQCD is applied at a density of $10n_{\rm s}$ and when incorporating additional transport upper bound of $c_s^2\leq0.781$, respectively.}
    \label{fig:mr-tov}
    \hfill
\end{figure}
The maximum mass ($M_{\rm TOV}$) and corresponding radius ($R_{\rm TOV}$) of a nonrotating NS have been demonstrated to be crucial in probing the high-density EOS \citep{2024PhRvD.109h3037T}. We find that there is a remarkable consistency in these critical macroscopic properties of NSs (see Figure~\ref{fig:mr-tov}) when comparing results obtained with the transport bound to those constrained by pQCD, which also validates the $M_{\rm TOV}$ found in \citet{2024PhRvD.109d3052F}. Similarly, the mass-radius relationships also exhibit notable consistency, as detailed in the Appendix.

\begin{figure}
    \centering
    \includegraphics[width=0.49\textwidth]{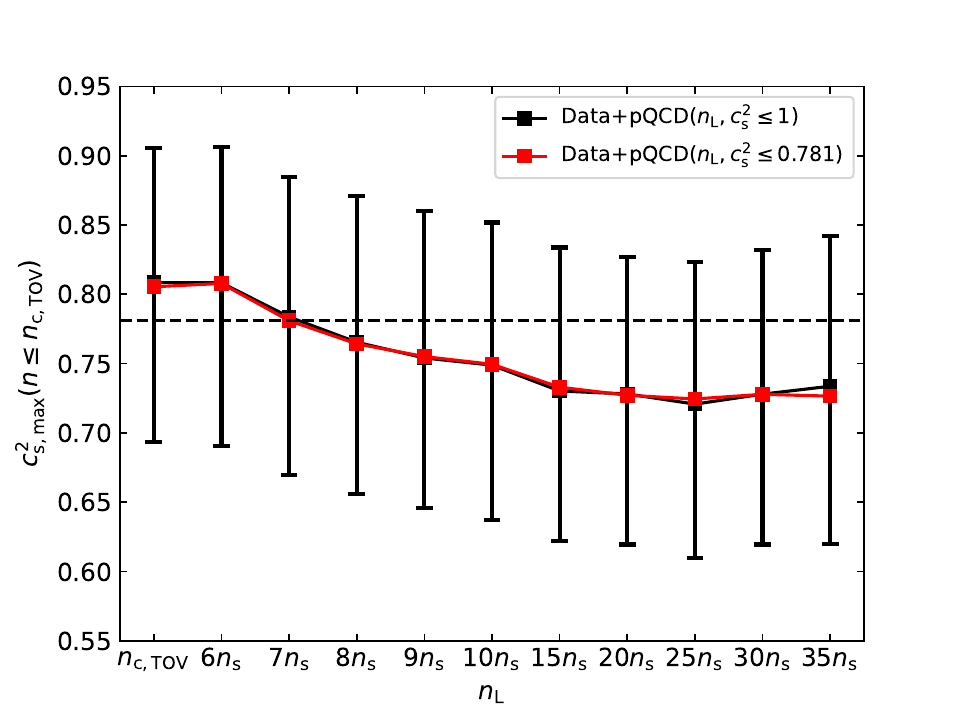}
    \caption{Correlation between $n_{\rm L}$ and the corresponding maximal values of $c_s^2$ within NS ($c_{\rm s, max}^2(n\leq n_{\rm c,TOV})$). The black $1\sigma$ error bar represents the results derived from extrapolating pQCD constraints to lower densities with causality. In contrast, the red line represents the results for the extension with $c_s^2 \leq 0.781$.}
    \label{fig:maxcs2-nqcd}
    \hfill
\end{figure}
It has been argued in the literature that the state of the massive NS core strongly depends on $n_{\rm L}$. Specifically, the quark matter core is (is not) required to arise in the massive NS, when  $n_{\rm L}=10n_{\rm s}$ ($n_{\rm L}=n_{\rm c,TOV}$) \citep{2023SciBu..68..913H, 2023NatCo..14.8451A}.
We have examined the `typical' densities at which pQCD likelihood is applied, including $n_{\rm c,TOV}$ and $10n_{\rm s}$. As depicted in Figure~\ref{fig:maxcs2-nqcd}, a negative correlation between $n_{\rm L}$ and $c_{\rm s, max}^2(n\leq n_{\rm c,TOV})$ is observed. With the increment of $n_{\rm L}$, the value of $c_{\rm s, max}^2(n\leq n_{\rm c,TOV})$ exhibits a gradual decrease until it approaches approximately $15n_{\rm s}$, beyond which it remains nearly constant. This intriguing pattern offers a novel approach to probing the appropriate densities for the application of pQCD likelihood. Given that such densities are not directly observable, the maximum values of $c_s^2$ within NSs can be indirectly constrained through astrophysical observations \citep{2022ApJ...939L..35E}. 
Our result suggests that $c_{\rm s, max}^2(n\leq n_{\rm c,TOV})$ given by higher $n_{\rm L}$ values aligns more closely with transport calculations, supporting the existence of a quark matter core in massive NS.

\section{Summary and Discussion} \label{sec:summary}
Prior research has established that the sound speed must surpass the conformal limit at certain densities to sustain observations of massive NSs \citep{2015PhRvL.114c1103B, 2020PhRvC.101d5803R, 2023SciBu..68..913H, 2024PhRvD.109h3037T}.
The speed of sound (as a function of density) within the NS EOS is believed to be nonmonotonic, characterized by the presence of at least one peak. The minimum of this peak is presumed to coincide with the conformal limit. A recent study by \citet{2024arXiv240214085H} proposes that the square of the sound speed is subject to an upper limit, derived from nuclear matter transport calculations, although this is still in debate \citep{2024JHEP...06..171M}. Determining whether there exists a more stringent or readily attainable upper limit than that imposed by causality in nature is a compelling avenue for further study.
In this study, the new radius measurements for PSR J0437$\textendash$4715 ($11.36^{+0.95}_{-0.63}$ km), PSR J0030+0451 ($11.71^{+0.88}_{-0.83}$ km, in the ST+PDT model), and PSR J0740+6620 ($12.49^{+1.28}_{-0.88}$ km) have been adopted. We find that multimessenger NS observations, when combined with pQCD likelihood applied at high densities (e.g., $n \geq 10n_{\rm s}$), lead to an upper bound on $c_s^2$, which is consistent with the bound from transport calculations. Additionally, we identify a negative correlation between $n_{\rm L}$ and the $c_{\rm s,max}^2$ that may become observable with future multimessenger observations. Therefore, it is more natural to consider a higher $n_{\rm L}$ for pQCD likelihood. We have examined the robustness of the above results against previous radius measurements for PSR J0030+0451 \citep{2019ApJ...887L..21R, 2019ApJ...887L..24M}. The new estimation for $R_{1.4}$ is smaller than the earlier reported value, resulting in a marginally higher maximum $c_s^2$ inside the NS (see also the Appendix of \citet{2024PhRvD.109d3052F}), while the consistency still holds. Moreover, if the upper bound by transport is assumed to be valid across the entire EOS, it also reproduces many predictions derived from pQCD likelihood, including the pressure-density and the NS mass-radius relationships. This suggests that these two distinctly different approaches—pQCD and transport calculations—converge on the same conclusion. Such convergence hints at a profound underlying similarity in the fundamental properties of nature.

\begin{acknowledgments}
This work is supported by the National Natural Science Foundation of China under Grants No. 12233011 and No. 12303056, the Project for Special Research Assistant and the Project for Young Scientists in Basic Research (No. YSBR-088) of the Chinese Academy of Sciences, the General Fund (No. 2023M733735, No. 2023M733736) of the China Postdoctoral Science Foundation (CPSF), and the Postdoctoral Fellowship Program of CPSF (GZB20230839, GZC20241915).
\end{acknowledgments}

\appendix

\begin{figure}
    \centering
    \includegraphics[width=0.49\textwidth]{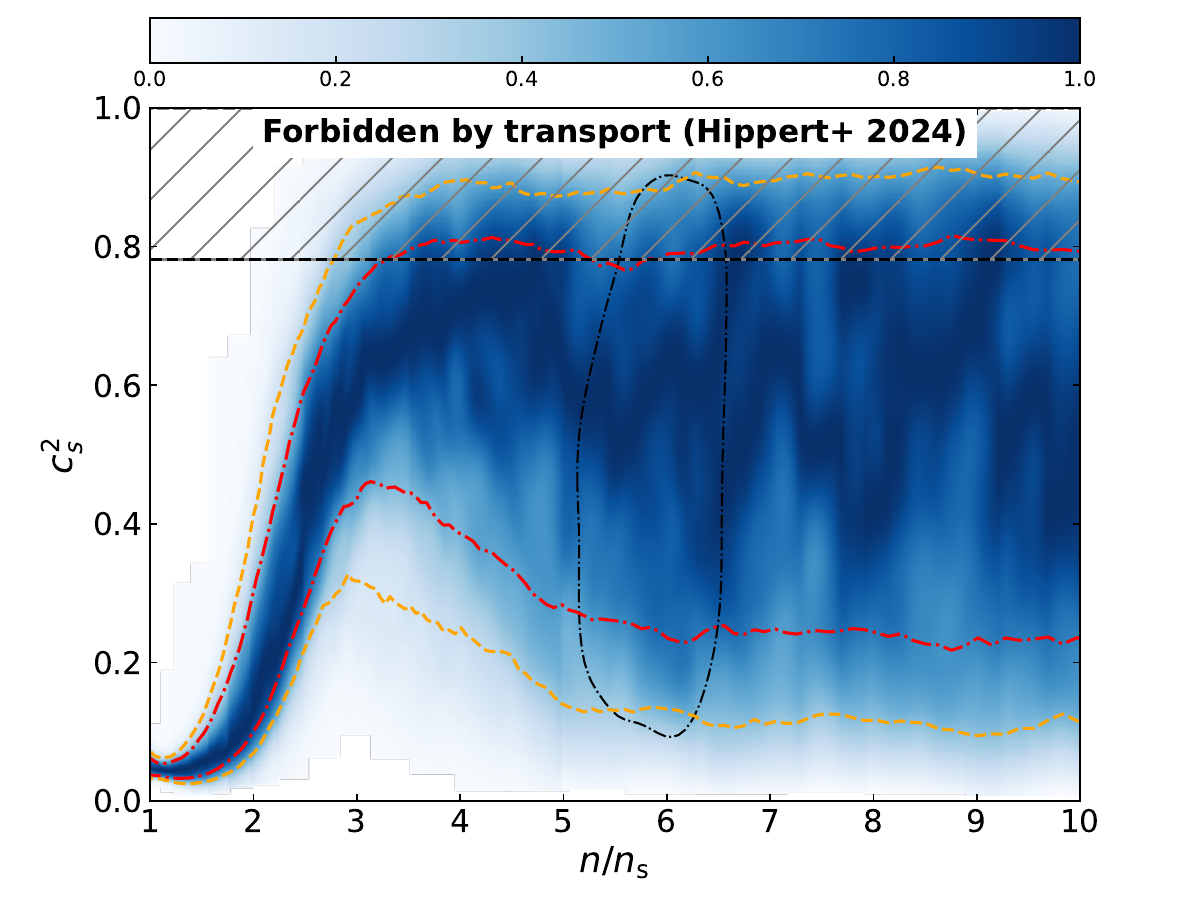}
    \includegraphics[width=0.49\textwidth]{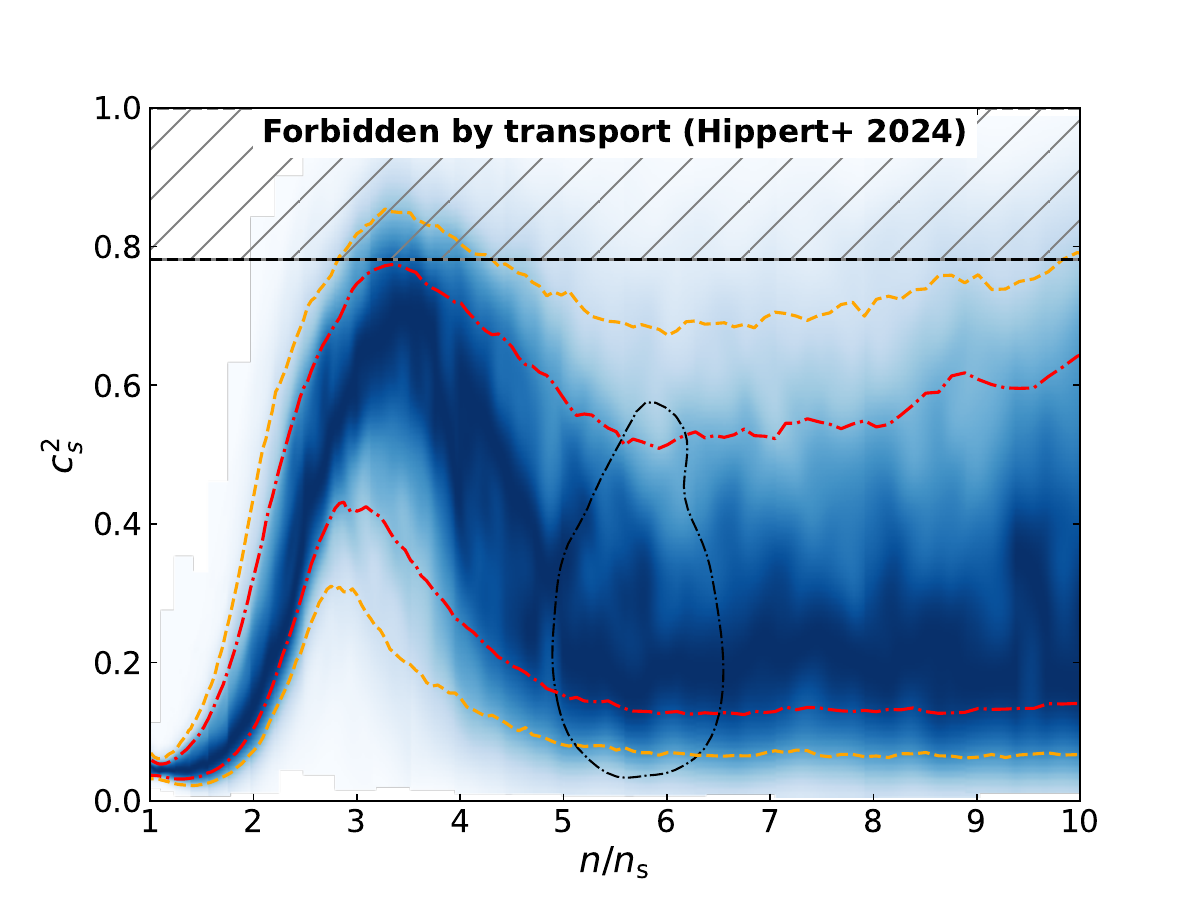}
    \caption{The same as Figure~1 but derived using $c_s^2\leq 0.781$ extension of pQCD to lower densities.}
    \label{fig:n-cs2-0781}
    \hfill
\end{figure}
In the main text, we have examined the effect of varying the density $n_{\rm L}$ at which the pQCD is applied. Specifically, we focus on the densities $n_{\rm L}=n_{\rm c,TOV}$ and $n_{\rm L}=10n_{\rm s}$. These variations significantly influence the structure of the $c_s^2-n$ relationship. The findings discussed are predicated on an extension of pQCD to lower densities that adheres to the principle of causality. Here we provide complementary results derived from extending pQCD to lower densities while adhering to the transport upper bound. As illustrated in Figure~\ref{fig:n-cs2-0781}, the outcomes are nearly identical to those previously reported.

\begin{figure}
    \centering
    \includegraphics[width=0.49\textwidth]{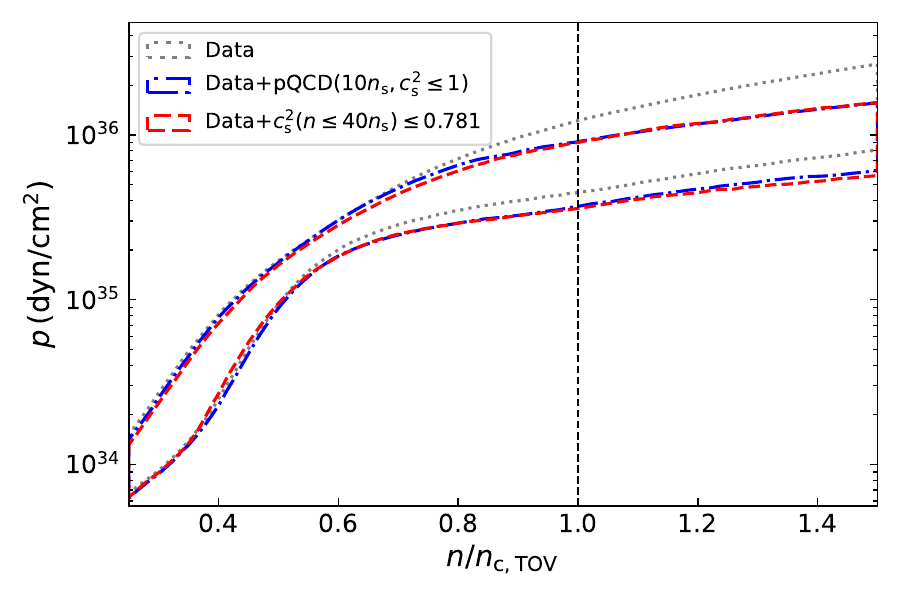}
    \caption{Similar to Figure~2 but for the pressure versus number density (normalized to $n_{\rm c,TOV}$) relationship.}
    \label{fig:nnctov-p}
    \hfill
\end{figure}
In the main text, we have presented a comparison of the pressure versus rest-mass density relationships derived from the imposition of a transport upper bound to $c_s^2$ and that obtained with the likelihood of pQCD applied at $n_{\rm L}=10n_{\rm s}$. Here we normalize the pressure versus number density results to the central density of a maximally massive nonrotating NS, $n_{\rm c,TOV}$. These normalized results (as shown in Figure~\ref{fig:nnctov-p}) also demonstrate consistent agreement across the different methodologies employed.
\begin{figure}
    \centering
    \includegraphics[width=0.49\textwidth]{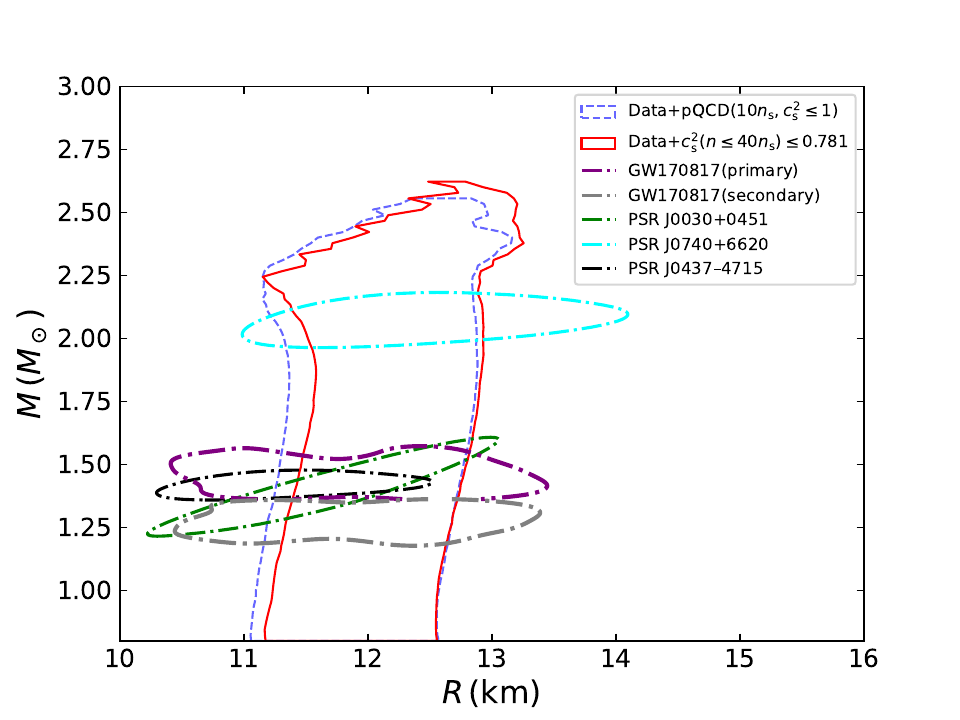}
    \caption{Reconstructed $90\%$ intervals of mass-radius curves. The blue and red lines depict the outcomes derived from multimessenger NS observations when combined with a pQCD likelihood applied at a density of $10n_{\rm s}$, and the outcomes obtained from multimessenger NS observations with the imposition of a transport upper bound on the speed of sound squared $c_s^2\leq0.781$, respectively. The green, cyan, purple, gray, and black dashed-dotted contours represent the $68.3\%$ mass-radius measurements for PSR J0030+0451 (i.e., the ST+PDT model results reported in \citet{2024ApJ...961...62V}), the very massive PSR J0740+6620 \citep{2024arXiv240614466S}, the primary NS of GW170817, the secondary component of GW170817 \citep{2018PhRvL.121p1101A}, and PSR J0437$\textendash$4715 \citep{2024ApJ...971L..20C}, respectively.}
    \label{fig:mr-range}
    \hfill
\end{figure}
We further examine the consistency of the mass-radius relationship. As depicted in Figure~\ref{fig:mr-range}, the findings from both approaches are in agreement and conform well to the observational data for NSs.

\begin{figure*}
    \centering
    \includegraphics[width=0.98\textwidth]{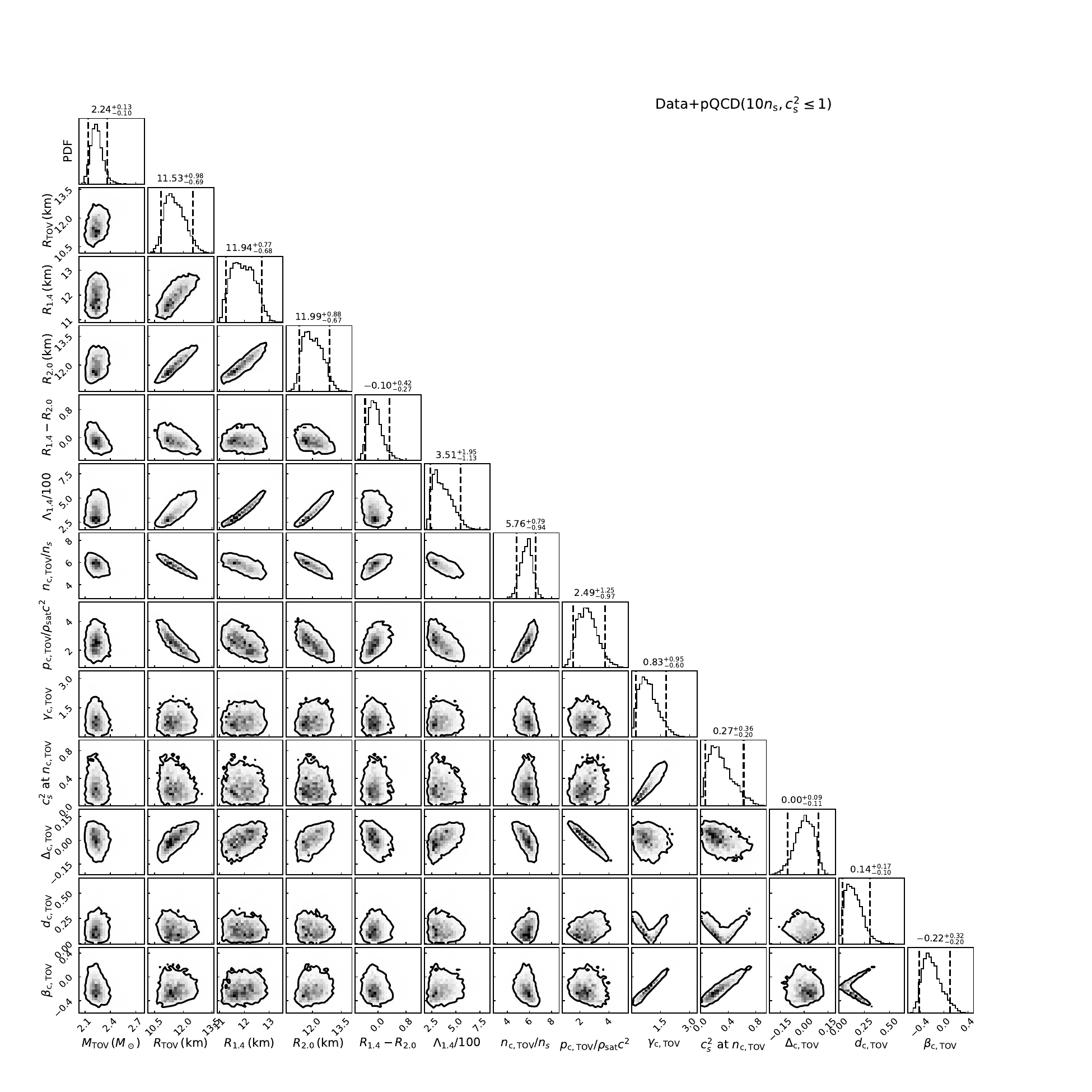}
    \caption{Corner plots of the probability distributions for the maximum mass ($M_{\rm TOV}$) and corresponding radius ($R_{\rm TOV}$) of a nonrotating NS, the radius ($R_{1.4}$) and tidal deformability ($\Lambda_{1.4}$) of a canonical $1.4M_\odot$ NS, the radius of a $2M_\odot$ NS ($R_{2.0}$), the difference between $R_{1.4}$ and $R_{2.0}$, and the number density ($n_{\rm c,TOV}$, normalized to nuclear saturation density $n_{\rm s}$), pressure ($p_{\rm c,TOV}$, normalized to the product of the rest-mass density at $n_{\rm s}$ and the square of the speed of light $c^2$), polytropic index ($\gamma_{\rm c,TOV}$), and the squared sound speed, and the normalized trace anomaly ($\Delta_{\rm c,TOV}$ \cite{2022PhRvL.129y2702F}) in the center of the NS with $M = M_{\rm TOV}$. Additionally, $d_{\rm c,TOV}$, a conformal criterion introduced by \citet{2023NatCo..14.8451A}, and $\beta_{\rm c,TOV}$, representing the curvature of the energy per particle as proposed by \citet{2024PhRvD.109d1302M}, are included. All uncertainties are reported at the $90\%$ credible level.}
    \label{fig:meta-corner}
    \hfill
\end{figure*}
Finally, we present the distributions of some key parameters. Figure~\ref{fig:meta-corner} depicts the probability distributions for the maximum mass ($M_{\rm TOV}$) and corresponding radius ($R_{\rm TOV}$) of a nonrotating NS, along with the radius ($R_{1.4}$) and tidal deformability ($\Lambda_{1.4}$) for a canonical $1.4M_\odot$ NS. 
The inferred $R_{1.4}$ and $\Lambda_{1.4}$ are constrained to $11.94_{-0.68}^{+0.77}$ km and $351_{-113}^{+195}$, which are strongly correlated with each other \citep{2018PhRvC..98c5804M} since these bulk properties are mainly determined by the pressure at approximately twice the nuclear saturation density ($2n_{\rm s}$) \citep{2016PhR...621..127L}.
The inferred radius of a $2M_\odot$ NS ($R_{2.0}$) is constrained to $11.99_{-0.67}^{+0.88}$ km, and the difference between $R_{1.4}$ and $R_{2.0}$ is constrained to $-0.1_{-0.27}^{+0.42}$ km.
We also investigate the central number density ($n_{\rm c,TOV}$), pressure ($p_{\rm c,TOV}$), polytropic index ($\gamma_{\rm c,TOV}$), and the squared speed of sound. These central properties are assessed for NSs at their maximum mass configuration ($M = M_{\rm TOV}$).
We notice that there is an intriguing anticorrelation between $R_{\rm TOV}$ and $n_{\rm c,TOV}$, which was also reported by \citet{2023ApJ...949...11J}.
Establishing the criterion for the emergence of strongly coupled conformal matter is crucial for determining the possible existence of a quark core within massive NSs.
For instance, the normalized trace anomaly ($\Delta_{\rm c,TOV}=1/3-p_{\rm c,TOV}/e_{\rm c,TOV}$), as discussed in \citet{2022PhRvL.129y2702F}, has been suggested as indicative of conformality in NSs.
Additionally, a conformal criterion, denoted as $d_{\rm c,TOV}=\sqrt{\Delta_{\rm c,TOV}^2+\Delta_{\rm c,TOV}^{\prime\,2}}$ was proposed by \citet{2023NatCo..14.8451A}.
In addition, \citet{2024PhRvD.109d1302M} argued that the curvature of the energy per particle $\beta_{\rm c,TOV}=c_{\rm s,TOV}^2-2(1/3-\Delta_{\rm c,TOV})/(4/3-\Delta_{\rm c,TOV})$ may serve as an approximate order parameter that signifies the onset of strongly coupled conformal matter in the NS core. All these parameters are included in the corner plots (i.e., Figure~\ref{fig:meta-corner}).

\bibliography{refs.bib}{}

\begin{thebibliography}{}
\expandafter\ifx\csname natexlab\endcsname\relax\def\natexlab#1{#1}\fi
\providecommand{\url}[1]{\href{#1}{#1}}
\providecommand{\dodoi}[1]{doi:~\href{http://doi.org/#1}{\nolinkurl{#1}}}
\providecommand{\doeprint}[1]{\href{http://ascl.net/#1}{\nolinkurl{http://ascl.net/#1}}}
\providecommand{\doarXiv}[1]{\href{https://arxiv.org/abs/#1}{\nolinkurl{https://arxiv.org/abs/#1}}}

\bibitem[{{Abbott} {et~al.}(2018){Abbott}, {Abbott}, {Abbott}, {Acernese},
  {Ackley}, {et~al.}}]{2018PhRvL.121p1101A}
{Abbott}, B.~P., {Abbott}, R., {Abbott}, T.~D., {et~al.} 2018, \prl, 121,
  161101, \dodoi{10.1103/PhysRevLett.121.161101}

\bibitem[{{Abbott} {et~al.}(2019){Abbott}, {Abbott}, {Abbott}, {Acernese},
  {Ackley}, {et~al.}}]{2019PhRvX...9a1001A}
---. 2019, Physical Review X, 9, 011001, \dodoi{10.1103/PhysRevX.9.011001}

\bibitem[{{Altiparmak} {et~al.}(2022){Altiparmak}, {Ecker}, \&
  {Rezzolla}}]{2022ApJ...939L..34A}
{Altiparmak}, S., {Ecker}, C., \& {Rezzolla}, L. 2022, \apjl, 939, L34,
  \dodoi{10.3847/2041-8213/ac9b2a}

\bibitem[{{Annala} {et~al.}(2023){Annala}, {Gorda}, {Hirvonen}, {Komoltsev},
  {Kurkela}, {N{\"a}ttil{\"a}}, \& {Vuorinen}}]{2023NatCo..14.8451A}
{Annala}, E., {Gorda}, T., {Hirvonen}, J., {et~al.} 2023, Nature
  Communications, 14, 8451, \dodoi{10.1038/s41467-023-44051-y}

\bibitem[{{Annala} {et~al.}(2020){Annala}, {Gorda}, {Kurkela},
  {N{\"a}ttil{\"a}}, \& {Vuorinen}}]{2020NatPh..16..907A}
{Annala}, E., {Gorda}, T., {Kurkela}, A., {N{\"a}ttil{\"a}}, J., \& {Vuorinen},
  A. 2020, Nature Physics, 16, 907, \dodoi{10.1038/s41567-020-0914-9}

\bibitem[{{Antoniadis} {et~al.}(2013){Antoniadis}, {Freire}, {Wex}, {Tauris},
  {Lynch}, {van Kerkwijk}, {Kramer}, {Bassa}, {et~al.}}]{2013Sci...340..448A}
{Antoniadis}, J., {Freire}, P. C.~C., {Wex}, N., {et~al.} 2013, Science, 340,
  448, \dodoi{10.1126/science.1233232}

\bibitem[{{Arnold} {et~al.}(2003){Arnold}, {Moore}, \&
  {Yaffe}}]{2003JHEP...05..051A}
{Arnold}, P., {Moore}, G.~D., \& {Yaffe}, L.~G. 2003, Journal of High Energy
  Physics, 2003, 051, \dodoi{10.1088/1126-6708/2003/05/051}

\bibitem[{{Baier} {et~al.}(2008){Baier}, {Romatschke}, {Thanh Son},
  {Starinets}, \& {Stephanov}}]{2008JHEP...04..100B}
{Baier}, R., {Romatschke}, P., {Thanh Son}, D., {Starinets}, A.~O., \&
  {Stephanov}, M.~A. 2008, Journal of High Energy Physics, 2008, 100,
  \dodoi{10.1088/1126-6708/2008/04/100}

\bibitem[{{Bedaque} \& {Steiner}(2015)}]{2015PhRvL.114c1103B}
{Bedaque}, P., \& {Steiner}, A.~W. 2015, \prl, 114, 031103,
  \dodoi{10.1103/PhysRevLett.114.031103}

\bibitem[{{Brandes} {et~al.}(2023{\natexlab{a}}){Brandes}, {Weise}, \&
  {Kaiser}}]{2023PhRvD.108i4014B}
{Brandes}, L., {Weise}, W., \& {Kaiser}, N. 2023{\natexlab{a}}, \prd, 108,
  094014, \dodoi{10.1103/PhysRevD.108.094014}

\bibitem[{{Brandes} {et~al.}(2023{\natexlab{b}}){Brandes}, {Weise}, \&
  {Kaiser}}]{2023PhRvD.107a4011B}
---. 2023{\natexlab{b}}, \prd, 107, 014011, \dodoi{10.1103/PhysRevD.107.014011}

\bibitem[{{Choudhury} {et~al.}(2024){Choudhury}, {Salmi}, {Vinciguerra},
  {Riley}, {Kini}, {Watts}, {Dorsman}, {Bogdanov}, {Guillot}, {Ray}, {Reardon},
  {Remillard}, {Bilous}, {Huppenkothen}, {Lattimer}, {Rutherford},
  {Arzoumanian}, {Gendreau}, {Morsink}, \& {Ho}}]{2024ApJ...971L..20C}
{Choudhury}, D., {Salmi}, T., {Vinciguerra}, S., {et~al.} 2024, \apjl, 971,
  L20, \dodoi{10.3847/2041-8213/ad5a6f}

\bibitem[{{Dittmann} {et~al.}(2024){Dittmann}, {Miller}, {Lamb}, {Holt},
  {Chirenti}, {Wolff}, {Bogdanov}, {Guillot}, {Ho}, {Morsink}, {Arzoumanian},
  \& {Gendreau}}]{2024arXiv240614467D}
{Dittmann}, A.~J., {Miller}, M.~C., {Lamb}, F.~K., {et~al.} 2024, arXiv
  e-prints, arXiv:2406.14467, \dodoi{10.48550/arXiv.2406.14467}

\bibitem[{{Ecker} \& {Rezzolla}(2022)}]{2022ApJ...939L..35E}
{Ecker}, C., \& {Rezzolla}, L. 2022, \apjl, 939, L35,
  \dodoi{10.3847/2041-8213/ac8674}

\bibitem[{{Essick} {et~al.}(2020){Essick}, {Landry}, \&
  {Holz}}]{2020PhRvD.101f3007E}
{Essick}, R., {Landry}, P., \& {Holz}, D.~E. 2020, \prd, 101, 063007,
  \dodoi{10.1103/PhysRevD.101.063007}

\bibitem[{{Essick} {et~al.}(2023){Essick}, {Legred}, {Chatziioannou}, {Han}, \&
  {Landry}}]{2023PhRvD.108d3013E}
{Essick}, R., {Legred}, I., {Chatziioannou}, K., {Han}, S., \& {Landry}, P.
  2023, \prd, 108, 043013, \dodoi{10.1103/PhysRevD.108.043013}

\bibitem[{{Fan} {et~al.}(2024){Fan}, {Han}, {Jiang}, {Shao}, \&
  {Tang}}]{2024PhRvD.109d3052F}
{Fan}, Y.-Z., {Han}, M.-Z., {Jiang}, J.-L., {Shao}, D.-S., \& {Tang}, S.-P.
  2024, \prd, 109, 043052, \dodoi{10.1103/PhysRevD.109.043052}

\bibitem[{{Fonseca} {et~al.}(2021){Fonseca}, {Cromartie}, {Pennucci}, {Ray},
  {Kirichenko}, {Ransom}, {Demorest}, {Stairs}, {et~al.}}]{2021ApJ...915L..12F}
{Fonseca}, E., {Cromartie}, H.~T., {Pennucci}, T.~T., {et~al.} 2021, \apjl,
  915, L12, \dodoi{10.3847/2041-8213/ac03b8}

\bibitem[{{Fujimoto} {et~al.}(2022){Fujimoto}, {Fukushima}, {McLerran}, \&
  {Prasza{\l}owicz}}]{2022PhRvL.129y2702F}
{Fujimoto}, Y., {Fukushima}, K., {McLerran}, L.~D., \& {Prasza{\l}owicz}, M.
  2022, \prl, 129, 252702, \dodoi{10.1103/PhysRevLett.129.252702}

\bibitem[{{Gorda} {et~al.}(2023){Gorda}, {Komoltsev}, \&
  {Kurkela}}]{2023ApJ...950..107G}
{Gorda}, T., {Komoltsev}, O., \& {Kurkela}, A. 2023, \apj, 950, 107,
  \dodoi{10.3847/1538-4357/acce3a}

\bibitem[{{Han} {et~al.}(2023){Han}, {Huang}, {Tang}, \&
  {Fan}}]{2023SciBu..68..913H}
{Han}, M.-Z., {Huang}, Y.-J., {Tang}, S.-P., \& {Fan}, Y.-Z. 2023, Science
  Bulletin, 68, 913, \dodoi{10.1016/j.scib.2023.04.007}

\bibitem[{{Hippert} {et~al.}(2024){Hippert}, {Noronha}, \&
  {Romatschke}}]{2024arXiv240214085H}
{Hippert}, M., {Noronha}, J., \& {Romatschke}, P. 2024, arXiv e-prints,
  arXiv:2402.14085, \dodoi{10.48550/arXiv.2402.14085}

\bibitem[{{Jiang} {et~al.}(2023){Jiang}, {Ecker}, \&
  {Rezzolla}}]{2023ApJ...949...11J}
{Jiang}, J.-L., {Ecker}, C., \& {Rezzolla}, L. 2023, \apj, 949, 11,
  \dodoi{10.3847/1538-4357/acc4be}

\bibitem[{{Komoltsev} \& {Kurkela}(2022)}]{2022PhRvL.128t2701K}
{Komoltsev}, O., \& {Kurkela}, A. 2022, \prl, 128, 202701,
  \dodoi{10.1103/PhysRevLett.128.202701}

\bibitem[{{Komoltsev} {et~al.}(2024){Komoltsev}, {Somasundaram}, {Gorda},
  {Kurkela}, {Margueron}, \& {Tews}}]{2024PhRvD.109i4030K}
{Komoltsev}, O., {Somasundaram}, R., {Gorda}, T., {et~al.} 2024, \prd, 109,
  094030, \dodoi{10.1103/PhysRevD.109.094030}

\bibitem[{{Kurkela}(2022)}]{2022arXiv221111414K}
{Kurkela}, A. 2022, arXiv e-prints, arXiv:2211.11414,
  \dodoi{10.48550/arXiv.2211.11414}

\bibitem[{{Landry} \& {Essick}(2019)}]{2019PhRvD..99h4049L}
{Landry}, P., \& {Essick}, R. 2019, \prd, 99, 084049,
  \dodoi{10.1103/PhysRevD.99.084049}

\bibitem[{{Landry} {et~al.}(2020){Landry}, {Essick}, \&
  {Chatziioannou}}]{2020PhRvD.101l3007L}
{Landry}, P., {Essick}, R., \& {Chatziioannou}, K. 2020, \prd, 101, 123007,
  \dodoi{10.1103/PhysRevD.101.123007}

\bibitem[{{Lattimer} \& {Prakash}(2016)}]{2016PhR...621..127L}
{Lattimer}, J.~M., \& {Prakash}, M. 2016, \physrep, 621, 127,
  \dodoi{10.1016/j.physrep.2015.12.005}

\bibitem[{{Luo} {et~al.}(2024){Luo}, {Tang}, {Han}, {Jiang}, {Gao}, \&
  {Wei}}]{2024ApJ...966...98L}
{Luo}, C.-N., {Tang}, S.-P., {Han}, M.-Z., {et~al.} 2024, \apj, 966, 98,
  \dodoi{10.3847/1538-4357/ad39ed}

\bibitem[{{Malik} {et~al.}(2018){Malik}, {Alam}, {Fortin}, {Provid{\^e}ncia},
  {Agrawal}, {Jha}, {Kumar}, \& {Patra}}]{2018PhRvC..98c5804M}
{Malik}, T., {Alam}, N., {Fortin}, M., {et~al.} 2018, \prc, 98, 035804,
  \dodoi{10.1103/PhysRevC.98.035804}

\bibitem[{{Marczenko} {et~al.}(2024){Marczenko}, {Redlich}, \&
  {Sasaki}}]{2024PhRvD.109d1302M}
{Marczenko}, M., {Redlich}, K., \& {Sasaki}, C. 2024, \prd, 109, L041302,
  \dodoi{10.1103/PhysRevD.109.L041302}

\bibitem[{{Miller} {et~al.}(2019){Miller}, {Lamb}, {Dittmann},
  {et~al.}}]{2019ApJ...887L..24M}
{Miller}, M.~C., {Lamb}, F.~K., {Dittmann}, A.~J., {et~al.} 2019, \apjl, 887,
  L24, \dodoi{10.3847/2041-8213/ab50c5}

\bibitem[{{Moore}(2024)}]{2024JHEP...06..171M}
{Moore}, G.~D. 2024, Journal of High Energy Physics, 2024, 171,
  \dodoi{10.1007/JHEP06(2024)171}

\bibitem[{{Moore} \& {Sohrabi}(2012)}]{2012JHEP...11..148M}
{Moore}, G.~D., \& {Sohrabi}, K.~A. 2012, Journal of High Energy Physics, 2012,
  148, \dodoi{10.1007/JHEP11(2012)148}

\bibitem[{{Mroczek} {et~al.}(2023){Mroczek}, {Miller}, {Noronha-Hostler}, \&
  {Yunes}}]{2023arXiv230902345M}
{Mroczek}, D., {Miller}, M.~C., {Noronha-Hostler}, J., \& {Yunes}, N. 2023,
  arXiv e-prints, arXiv:2309.02345, \dodoi{10.48550/arXiv.2309.02345}

\bibitem[{{Policastro} {et~al.}(2001){Policastro}, {Son}, \&
  {Starinets}}]{2001PhRvL..87h1601P}
{Policastro}, G., {Son}, D.~T., \& {Starinets}, A.~O. 2001, \prl, 87, 081601,
  \dodoi{10.1103/PhysRevLett.87.081601}

\bibitem[{{Reed} \& {Horowitz}(2020)}]{2020PhRvC.101d5803R}
{Reed}, B., \& {Horowitz}, C.~J. 2020, \prc, 101, 045803,
  \dodoi{10.1103/PhysRevC.101.045803}

\bibitem[{{Riley} {et~al.}(2019){Riley}, {Watts}, {Bogdanov},
  {et~al.}}]{2019ApJ...887L..21R}
{Riley}, T.~E., {Watts}, A.~L., {Bogdanov}, S., {et~al.} 2019, \apjl, 887, L21,
  \dodoi{10.3847/2041-8213/ab481c}

\bibitem[{{Romatschke} \& {Romatschke}(2007)}]{2007PhRvL..99q2301R}
{Romatschke}, P., \& {Romatschke}, U. 2007, \prl, 99, 172301,
  \dodoi{10.1103/PhysRevLett.99.172301}

\bibitem[{{Romatschke} \& {Romatschke}(2017)}]{2017arXiv171205815R}
---. 2017, arXiv e-prints, arXiv:1712.05815, \dodoi{10.48550/arXiv.1712.05815}

\bibitem[{{Salmi} {et~al.}(2024){Salmi}, {Choudhury}, {Kini}, {Riley},
  {Vinciguerra}, {Watts}, {Wolff}, {Arzoumanian}, {Bogdanov}, {Chakrabarty},
  {Gendreau}, {Guillot}, {Ho}, {Huppenkothen}, {Ludlam}, {Morsink}, \&
  {Ray}}]{2024arXiv240614466S}
{Salmi}, T., {Choudhury}, D., {Kini}, Y., {et~al.} 2024, arXiv e-prints,
  arXiv:2406.14466, \dodoi{10.48550/arXiv.2406.14466}

\bibitem[{{Somasundaram} {et~al.}(2023){Somasundaram}, {Tews}, \&
  {Margueron}}]{2023PhRvC.107e2801S}
{Somasundaram}, R., {Tews}, I., \& {Margueron}, J. 2023, \prc, 107, L052801,
  \dodoi{10.1103/PhysRevC.107.L052801}

\bibitem[{{Tang} {et~al.}(2024){Tang}, {Han}, {Huang}, {Fan}, \&
  {Wei}}]{2024PhRvD.109h3037T}
{Tang}, S.-P., {Han}, M.-Z., {Huang}, Y.-J., {Fan}, Y.-Z., \& {Wei}, D.-M.
  2024, \prd, 109, 083037, \dodoi{10.1103/PhysRevD.109.083037}

\bibitem[{{Vinciguerra} {et~al.}(2024){Vinciguerra}, {Salmi}, {Watts},
  {Choudhury}, {Riley}, {Ray}, {Bogdanov}, {Kini}, {Guillot}, {Chakrabarty},
  {Ho}, {Huppenkothen}, {Morsink}, {Wadiasingh}, \&
  {Wolff}}]{2024ApJ...961...62V}
{Vinciguerra}, S., {Salmi}, T., {Watts}, A.~L., {et~al.} 2024, \apj, 961, 62,
  \dodoi{10.3847/1538-4357/acfb83}

\bibitem[{{Vuorinen}(2024)}]{2024AcPPB..55....1V}
{Vuorinen}, A. 2024, Acta Physica Polonica B, 55, 1,
  \dodoi{10.5506/APhysPolB.55.4-A4}

\bibitem[{{Yao} {et~al.}(2024){Yao}, {Sorensen}, {Dexheimer}, \&
  {Noronha-Hostler}}]{2024PhRvC.109f5803Y}
{Yao}, N., {Sorensen}, A., {Dexheimer}, V., \& {Noronha-Hostler}, J. 2024,
  \prc, 109, 065803, \dodoi{10.1103/PhysRevC.109.065803}

\bibitem[{{Zhou}(2023)}]{2023arXiv230711125Z}
{Zhou}, D. 2023, arXiv e-prints, arXiv:2307.11125,
  \dodoi{10.48550/arXiv.2307.11125}

\end{thebibliography}
\bibliographystyle{aasjournal}

\end{document}